\documentclass[acmsmall]{acmart}

\usepackage{booktabs}
\usepackage{graphicx}

\setcopyright{rightsretained}
\acmJournal{PACMHCI}
\acmYear{2021} 
\acmVolume{5} 
\acmNumber{CSCW2} 
\acmArticle{292} 
\acmMonth{10} 
\acmPrice{}
\acmDOI{10.1145/3476023}

\begin{document}

\title{Social Catalysts: Characterizing People Who Spark Conversations Among Others}

\author{Martin Saveski}
\email{msaveski@mit.edu}
\affiliation{%
  \institution{Massachusetts Institute of Technology}
}

\author{Farshad Kooti}
\email{farshadkt@fb.com}
\affiliation{%
  \institution{Facebook}
}

\author{Sylvia Morelli Vitousek}
\email{sylviav@fb.com}
\affiliation{%
  \institution{Facebook}
}

\author{Carlos Diuk}
\email{cdiuk@fb.com}
\affiliation{%
  \institution{Facebook}
}

\author{Bryce Bartlett}
\email{bjb@fb.com}
\affiliation{%
  \institution{Facebook}
}

\author{Lada Adamic}
\email{ladamic@fb.com}
\affiliation{%
  \institution{Facebook}
}

\renewcommand{\shortauthors}{Saveski et al.}

\begin{abstract}
People assume different and important roles within social networks. Some roles have received extensive study: that of influencers who are well-connected, and that of brokers who bridge unconnected parts of the network. However, very little work has explored another potentially important role, that of creating opportunities for people to interact and facilitating conversation between them. These individuals bring people together and act as \textit{social catalysts}.
In this paper, we test for the presence of social catalysts on the online social network Facebook. We first identify posts that have spurred conversations between the poster's friends and summarize the characteristics of such posts. We then aggregate the number of catalyzed comments at the poster level, as a measure of the individual's ``catalystness.'' 
The top 1\% of such individuals account for 31\% of catalyzed interactions, although their network characteristics do not differ markedly from others who post as frequently and have a similar number of friends.
By collecting survey data, we also validate the behavioral measure of catalystness: a person is more likely to be nominated as a social catalyst by their friends if their posts prompt discussions between other people more frequently. The measure, along with other conversation-related features, is one of the most predictive of a person being nominated as a catalyst. Although influencers and brokers may have gotten more attention for their network positions, our findings provide converging evidence that another important role exists and is recognized in online social networks.
\end{abstract}

%
%


%
%

\maketitle

\section{Introduction}
People can play different roles in social networks. Some are influencers, shaping how many people in their network behave via their numerous and important ties ~\cite{katz1957two,katz1955personal,watts2007influentials,banerjee2013diffusion,bakshy2011everyone,aral2012identifying}. Others are brokers, bringing value by transmitting ideas and information between disconnected communities ~\cite{granovetter1973strength,burt2005brokerage,burt2009structural,burt2017structural,uzzi1997social,gonzalez2013bridges}. While the behaviors and characteristics of influencers and brokers have been studied extensively, much less work has investigated individuals who bring other people together. We propose that \textit{social catalysts} create opportunities for people to interact and facilitate conversation between them.

In this paper, we seek to understand the role of social catalysts on Facebook. We introduce a measure of social catalystness: the extent to which an action by one individual (e.g., posting a status update) generates conversation between that person's friends (Section~\ref{sec:post-catalystness}). We then aggregate catalystness at the person level, and find a substantial skew in the amount of conversation individuals are prompting. Roughly 1\% of individuals prompt a third of the interactions between people on others' posts.

Seeking to validate the behavioral measures, we conducted a survey asking individuals to nominate up to three people in their network whose posts prompt interactions (Section~\ref{sec:survey}). The nominees had indeed generated a higher proportion of comment interactions, even after controlling for the number of posts and friends. Moreover, in a predictive model of which friend will be nominated as a catalyst, comment replies among one's friends is among the most predictive features.

In this work, we take the first steps in understanding social catalysts---people who are exceptionally good at bringing people together and facilitating interactions among others. Catalysts likely play a key role in enriching the social fabric, helping people integrate into new communities, enabling the creation of new ties, and strengthening the existing ones. Online, social catalysts may promote the kind of active engagement that has been associated with increased well-being~\cite{burke2016wellbeing}, and thus being able to identify catalyst activity may be useful in ranking and recommendations. More specifically, our approach could be used in any platform that has a commenting and replying functionality, such as Twitter, Youtube, and Reddit, to identify and promote catalyst users and content to foster more discussions among people.

In summary, we make the following contributions:
\begin{itemize}
\item We define a behavioral measure of catalystness based on the interactions prompted by the user's posts (Section~\ref{sec:measuring-catalystness}).
\item We use the behavioral measure to study the characteristics of catalysts posts. We find that they tend to be longer, original posts on more positive topics (Section~\ref{sec:post-catalystness}).
\item We analyze the characteristics of catalyst users and, surprisingly, find only small differences in the structure of the catalysts' social networks (Section~\ref{sec:user_cat}).
\item We run a survey asking users to nominate catalysts among their friends and find that our behavioral measure aligns with users' perceptions of catalystness (Section~\ref{sec:survey}). We also show that respondents' nominations can be accurately predicted (AUC = 78.58\%) using a machine learning model and that our behavioral measure is one of the most predictive features (Section~\ref{modeling-catalystness}).
\end{itemize}

In Section~\ref{sec:related_work}, we discuss how our paper relates to previous work in this area and in Section~\ref{sec:discussion_conclusion} we discuss the limitations of our work and propose directions for future research.\footnote{The code needed to replicate our analyses is available at: \url{https://github.com/facebookresearch/social-catalysts}.}

\section{Related Work \label{sec:related_work}}
Our investigation of social catalysts builds on previous work from several different research areas. In this section, we highlight the most related studies on social roles, specific social roles online, opinion leaders, and social capital. 

\subsection{Social roles}
Social roles have long been studied in the social sciences to understand an individual's behavior in organizations. The traditional social science literature has paid most attention to formal roles in organized structures, where roles are assigned and have well-defined expectations and responsibilities~\cite{newcomb1950role, biddle1986recent, klapp1958social}. In contrast, more recent work has focused on social roles in online communities, where roles are emergent, self-selected, and not formalized. In this context, social roles are often defined in terms of consistent and recurrent participation in certain activities or exhibition of certain behaviors~\cite{turner2005picturing, welser2007visualizing, gleave2009conceptual, sun2019multi, yang2019seekers}. 

One of the main benefits of defining social roles is that it reduces the complexity of social systems by grouping similar social relationships and behaviors into a smaller set of roles. This facilitates the comparative study of populations across different times and contexts~\cite{gleave2009conceptual}. In the context of online communities, social roles can be particularly beneficial in designing and managing social media spaces.

There are two main methodological approaches to studying social roles: \textit{interpretive} and \textit{structural}. Interpretative approaches rely on ethnographic methods~\cite{golder2004social}, content analysis~\cite{turner2006impact}, and surveys~\cite{brush2005assessing} to capture relationships and behavioral patterns within a group. On the other hand, structural approaches identify structural signatures that distinguish different group members by analyzing the networks of their relationships and interaction patterns~\cite{fisher2006you, welser2007visualizing, welser2011finding}. Interpretative approaches often lead to a deeper understanding of social roles and the context in which they emerge, but often miss the macro social structure in which these roles exist. Structural approaches focus on the macro social structures but abstract out the content and the context of relations. 

In this paper, we define the role of a social catalyst through the behavior of posting content that sparks conversations among others. However, to understand the role, we analyze the content of the posts that lead to catalyst interactions, the structure of the social networks that the catalysts are embedded in, and survey the friends of the catalysts to verify that our measure of catalystness matches with users' perceptions. 

We also note that many data mining methods for automated \textit{role discovery} have been recently proposed in the computer science literature~\cite{mccallum2007topic, henderson2012rolx, rossi2012role}. Role discovery is defined as the process that divides the nodes of a graph into sets of nodes with similar structural patterns. Some of these approaches operate on the graph, while others on features that describe the node positions. Rossi and Ahmed~\cite{rossi2014role} provide an extensive review of different methods for role discovery. In this work, our goal is to study in depth the characteristics of a single social role rather than to exhaustively determine the roles of all users on the platform.

\subsection{Social roles online}
Previous studies have investigated social roles in a wide variety of online spaces. Here, we discuss four social roles found in discussion forums that are most closely related to the current study. We refer the interested reader to~\cite{forestier2012roles} for a comprehensive survey of social roles online.

Several previous works have identified the role of an \textit{answer person}, a user whose dominant behavior is responding to questions posed by others~\cite{turner2005picturing, welser2007visualizing, buntain2014identifying}. These users typically reply to many different users and post only a few replies per thread. Answer people are critical to the forums in which they participate as they provide the value that attracts other users to these digital spaces. 

Another important role is that of a \textit{discussion person}~\cite{welser2007visualizing}, \textit{conversationalist}~\cite{turner2005picturing}, or a \textit{debater}~\cite{viegas2004newsgroup}. These users mostly engage in lengthy reciprocal exchanges with many other users, usually on conversation threads started by others. Their primary motivation for participating in discussion forums is communicating with others and evaluating ideas. As a result, they generate many valuable social interactions and bring a sense of belonging and community for other members. 

Graham and Wright~\cite{graham2014discursive} studied discussions in an online financial advice forum and identified \textit{superposters}, a minority of users who post disproportionately more often than other users. They found that these users constitute only 0.4\% of the user population but account for 47\% of the posts/replies on the forum. Superposters often shared personal stories and offered their expert advice. Although highly-active, they rarely attacked others or stopped them from engaging in conversations. In Section~\ref{sec:survey}, we investigate whether social catalysts exhibit the characteristics of superposters. We ask a sample of Facebook users to name social catalysts among their friends and compare the posting frequency of the nominated users with that of their other friends. 

Himelboim \textit{et al.}~\cite{himelboim2009discussion} identified the role of a \textit{discussion catalyst}, a user whose posts receive many replies. They studied political Usenet newsgroups and identified a small subset of users who start discussion threads that attract a disproportionately large number of replies. Similar to superposters, discussion catalysts post frequently, but also attract a large volume of responses. They find that these users post content published elsewhere, mainly news websites, and act as~filters~and~amplifiers. 

The role that we study in this paper, a social catalyst, is most related to that of a discussion catalyst. The main difference between the two is that social catalysts are not only successful at attracting many replies, but specifically attract interactions among others. This distinction is particularly important in the context of modern social media platforms like Facebook and Twitter. In contrast to public forums that resemble informal town meetings, users on these platforms have user profiles and the discussions prompted by their posts are clearly associated with them. As a result, most post comments and replies are discussions with the poster. However, sometimes posts stimulate interactions among other users. Our goal in this paper is to identify which post and user characteristics are associated with such interactions. As we show in later sections, social catalysts and discussion catalysts are two distinct sets of users. 

\subsection{Opinion leaders and influencers}
Katz and Lazarsfeld~\cite{katz1955personal, lazarsfeld1968peoples, katz1957two} proposed the most dominant theory of how the media influences people's views. They suggest that information flows through a few highly connected individuals, whom they call \textit{opinion leaders}, who first consume the information published by the media and then spread it to the rest of the population. Social catalysts may spark conversations among their friends by being opinion leaders and sharing their expert views or by bringing novel information into their networks. In Section~\ref{sec:post-catalystness}, we investigate the characteristics of catalyst posts and test whether they are more likely to be original posts vs. reshares or to contain links to external content.

The existence of disproportionately influential individuals inspired a large body of research on how to identify them. The promise of this research is that one can quickly reach and influence the entire population by influencing the most influential individuals first. While Katz and Lazarsfeld discovered the existence of opinion leaders using survey methods, most modern methods for identifying opinion leaders or influencers rely on social network analysis~\cite{easley2010networks, wasserman1994social, newman2010networks}. There are various methods for identifying a set of individuals that can spread information through the network and reach as many other people as possible and various metrics for quantifying the centrality of individuals in the network~\cite{kempe2003maximizing, banerjee2013diffusion}. One common heuristic for identifying influential users is by considering their degree~\cite{kimura2006tractable,chen2009efficient, kim2015social}. In Section~\ref{sec:survey}, we test whether users nominated by a survey participant have more Facebook friends than the participant's other friends. 

\subsection{Social capital}
Social capital refers to advantages people have because of how they are connected to others; which is a contextual complement to human capital, i.e., people's own skills and abilities~\cite{portes1998social, adler2002social}. There are two dominating theories of the network mechanisms that create social capital: closure and brokerage. 

According to Coleman~\cite{coleman1988social}, closure (network density) is the key facilitator of the creation of social capital. If everyone knows everyone else, then ($i$) people are more likely to trust each other and ($ii$) information flows faster. In contrast, Burt~\cite{burt2005brokerage, burt2009structural, burt2017structural} argues that structural holes—sparsity of connections among parts of the network—create social capital for the individuals whose ties span the structural holes. In the context of work settings, the ability to broker the flow of information is associated with higher productivity and creativity. 

This prior work has paid most attention to the information advantage brokers enjoy, but not necessarily whether they will catalyze interactions between different parts of the network.  
Similarly, much work has also been done in predicting when and where closure will occur~\cite{lu2011link,liben2007link,aiello2012friendship}. However, the role of individuals in initiating closure and even more so enabling continued interaction within a network that strengthens bonds, has not been studied.

In the context of social catalysts, each of these two mechanisms, closure and brokerage, can be at play. A person may be more likely to prompt conversations among her friends if many of her friends know and often communicate with each other. Alternatively, a person may spark interesting conversations among friends in one part of their network by brokering information that they discovered in another part of their network. In Section~\ref{subsec:network-structure}, we test whether social catalysts have markedly different social networks and whether the structure of their networks explains their behavior; and in Section~\ref{subsec:friendship-initiations}, we test whether social catalysts are more likely to connect friends and facilitate the closure of network triads.

\section{Data Description}
To understand the characteristics of social catalysts and catalyst posts, we study activity based on two sources: user behavior on Facebook and an on-platform survey.

To measure post catalystness (Section~\ref{sec:post-catalystness}) and behavioral measures of such activity aggregated to the user level (Section~\ref{sec:user_cat}), we analyzed all Facebook posts in the US shared publicly or with all the poster's friends during a 28-day period starting July 27, 2019. There are many mechanisms through which users can catalyze interactions on Facebook, e.g., by posting on their profile, posting on groups or pages, or in private messages. As this is our first attempt at understanding the characteristics and behaviors of social catalysts, we focus on the way users catalyze interactions among others through their posts. To exclude potentially confounding factors, we do not consider ($i$) posts in groups or pages where the engagement with the posts may be determined by the norms and dynamics of the group/page rather than the post's or the user's characteristics, and ($ii$) posts that tag other users, as it is unclear whether the poster or the tagged user(s) are responsible for catalyzing interactions.

To investigate whether the behavioral measure we developed corresponds with people's perceptions of social catalystness, we used 10,351 survey responses wherein people nominated up to three of their friends as facilitators of conversations among others. This analysis is presented~in~Section~\ref{sec:survey}.

To protect users' privacy, we performed all analyses on de-identified data, in aggregate, and following Facebook's Data Policy~\cite{facebookpolicy}. The research plan passed a rigorous internal review by Facebook to ensure that data is handled ethically and that users' privacy is preserved.

\section{Measuring Catalystness}
\label{sec:measuring-catalystness}

\begin{figure}[t]
\centering
\includegraphics[width=0.65\linewidth]{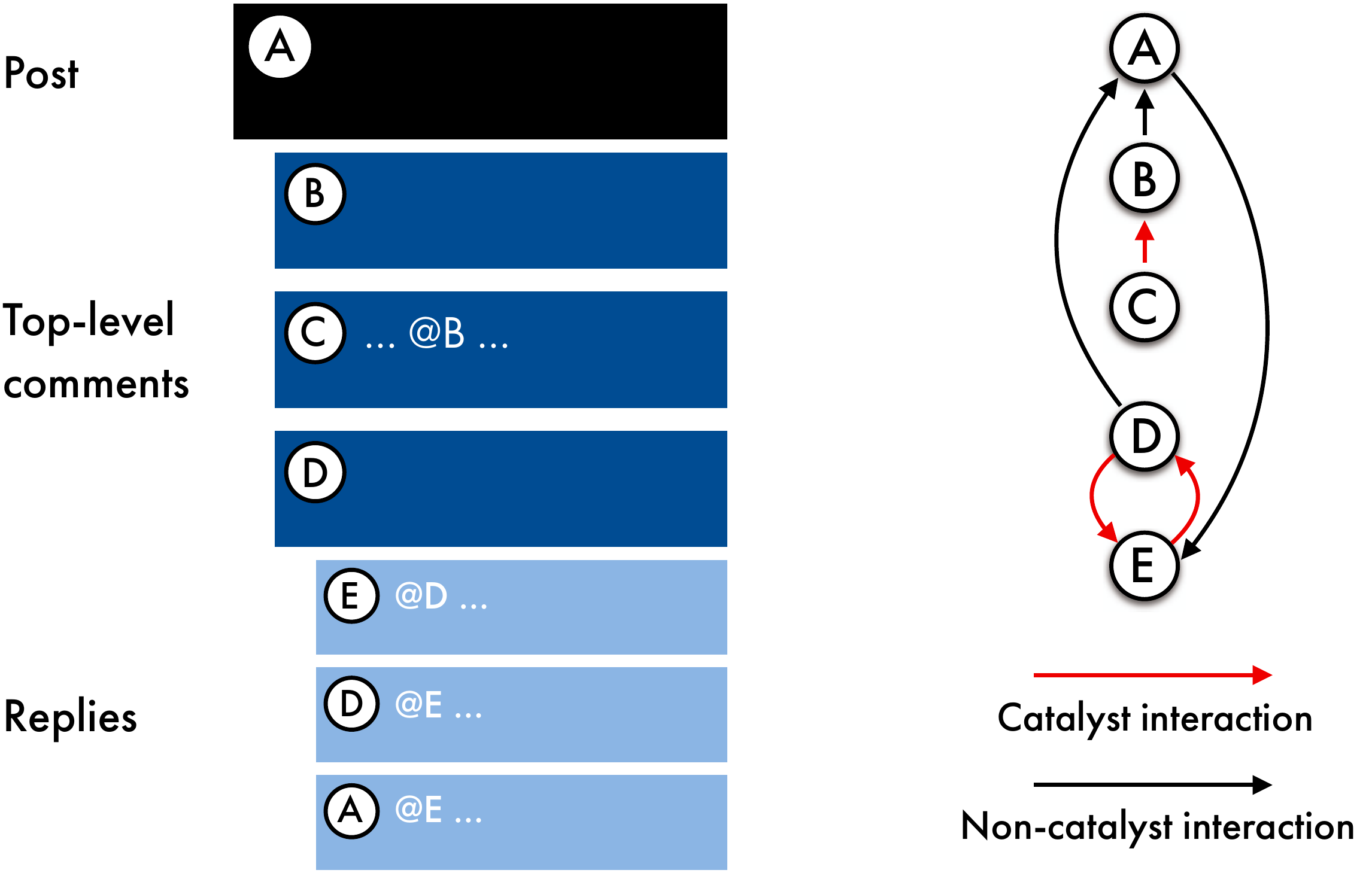}
\caption{
(Left) A sketch of the comment/reply structure on Facebook. User A posts a post; users B, C, and D comment; user E, D, and A reply to D's comment. Users can use @mentions to invoke another user in their comments/replies. (Right) The corresponding reply graph following the @mentions and the implicit structure of the comments/replies. The post has 3 catalyst replies: C$\rightarrow$B, E$\rightarrow$D, and D$\rightarrow$E (highlighted in red), resulting in a catalystness score of 3.}
\label{fig:comment_structure}
\end{figure}

As explained above, we consider social catalysts to be individuals who bring people together and prompt them to engage with one another. To identify such activity on Facebook, we consider the action of posting a status update, and the resulting exchange of comments and replies on the post. We define the catalystness measure for each post as the number of reply interactions among people {\em who are not the author of the post}. Specifically, if user A posts a status update, user B comments on it, and user C replies to B's comment; this reply is an interaction that user A has catalyzed.

To measure the catalystness for each post, we need to attribute to whom a given comment or reply is a response to. This way, we can exclude interactions with the post's author. Facebook's two-level comment/reply structure, shown in  Figure~\ref{fig:comment_structure}, presents some ambiguity about the intended recipient of each reply, since replies to replies are shown at the same level, and some top-level comments are intended as replies to earlier comments. Thus, to match replies to comments, we relied on @mentions of individuals in the thread. We identify a top-level comment as a reply if it includes a mention of another user who had commented on the thread earlier (e.g., Figure~\ref{fig:comment_structure}, the comment posted by user C mentioning user B). Similarly, in a thread of replies, we attribute the reply to the user mentioned. We note that when a user presses the ``reply'' button on a reply within a reply thread, the user interface automatically inserts a mention in the comment text box. In the absence of a mention, we attributed the reply to the comment the user pressed the ``reply'' button on, which can be a top-level comment or another reply. In all cases, if there is more than one mention matching these criteria, we attribute the interaction to the first mention. Using these rules together allows us to capture all interactions between the users.

In short, we define the catalystness score of a post as the number of comment-reply pairs among other users, and catalystness of a user as simply the sum of catalystness of posts created by that~user.

\section{Post Catalystness}
\label{sec:post-catalystness}
In this section, we compute the catalystness score for a large sample of posts and compare posts with high catalystness to posts with low catalystness in terms of their post type, length, and topics.

We analyze English comments on 120M posts posted by users in the US during a four-week period, starting June 27, 2019. First, we compute the catalystness score of every post, then we define catalyst posts as posts that have at least 15 catalyst interactions, which corresponds to roughly the top 1\% of all posts. The trends in findings hold with different thresholds; lower thresholds would make the differences smaller, and higher thresholds would make the differences even stronger. Due to the heavy-tailed distribution of engagement with posts, if we compare catalyst posts with a random sample of other posts, most of the differences could be simply explained by the audience size of the poster. Thus, to control for the poster's audience size, we match each catalyst post to another non-catalyst post posted by any user with the same number of Facebook friends\footnote{A matched (non-catalyst) post could be posted by a user who is a catalyst in general (as defined in Section~\ref{sec:user_cat}) but posted a non-catalyst post this time. We repeated the analysis in this section at the user level comparing the aggregate characteristics of posts by catalysts and matched, non-catalyst (see Section~\ref{sec:user_cat}) users and found very similar patterns.}. We use friend count as a proxy for the number of viewers the post will eventually reach.
In the rest of this section, we compare catalyst posts with matched posts posted by users with the same friend count as the author of the post being matched.

\subsection{Posts characteristics}
To understand how posts which catalyze discussion differ from others, we first compute general characteristics. Compared to matched posts, catalyst posts are 2.16 times (95\% CIs [2.14, 2.18]) more likely to be original posts and not reshares (catalysts: 75\%, matched: 23.7\% original posts). In both groups, photo posts are most common, followed by text, video, and link posts. Catalyst posts, however, are relatively less likely to be photo (-3.3\% difference, 95\% CIs [-3.48\%, -3.13\%]) and video (-13.4\% difference, 95\% CIs [-13.7\%, -13.2\%]) posts, and more likely to be text (+16.1\% difference, 95\% CIs [15.8\%, 16.3\%]) and link (+0.68\% difference,95\% CIs [0.44\%, 0.92\%]) posts.

Out of the posts that include text, catalyst posts tend to be longer, having 1.82 times (95\% CIs [1.76, 1.89]) more characters and 1.74 times (95\% CIs [1.68, 1.80]) more words on average, suggesting that they might be more informative. Also, catalyst posts, on average, contain 9.1\% (95\% CIs [3.28\%, 14.8\%]) more hashtags, and 11\% (95\% CIs [8.61\%, 13.5\%]) more emojis. Finally, catalyst posts are 2 times (95\% CIs [1.96, 2.04]) more likely to contain at least one question mark in their text, suggesting that they are more likely to pose questions which may serve as conversation prompts. 

The fact that catalyst posts are less likely to be reshares and only slightly more likely to be link posts suggests that social catalysts are different from opinion leaders~\cite{katz1955personal, lazarsfeld1968peoples} and discussion catalysts~\cite{himelboim2009discussion}, and do not catalyze interactions by filtering and amplifying content published~by~others.

\begin{figure}[t]
\centering
\includegraphics[width=0.85\linewidth]{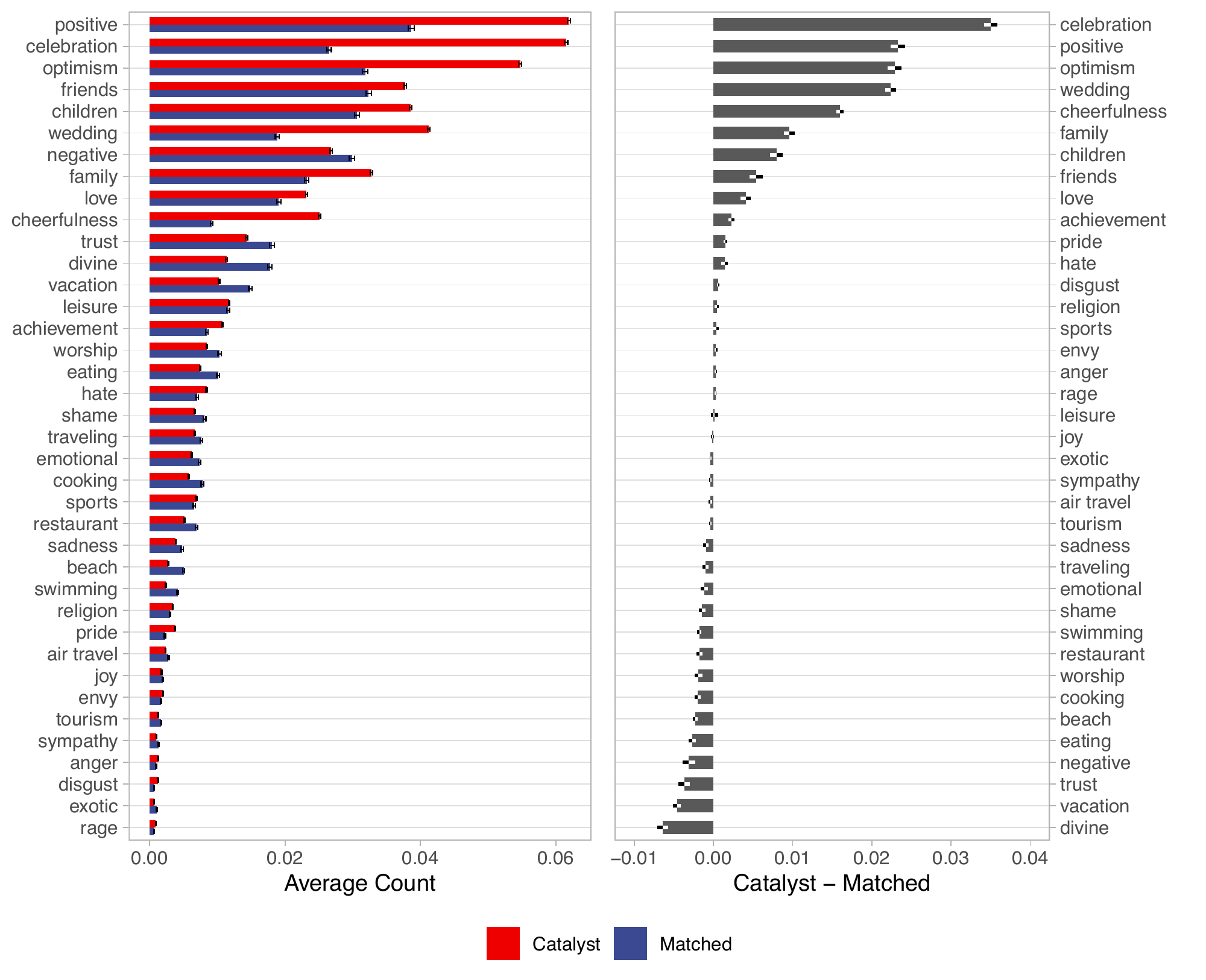}
\caption{Absolute (left) and relative (right) comparison between the topics found in catalysts vs. matched posts. The error bars represent 95\% confidence intervals.}
\label{fig:topics}
\end{figure}

\subsection{Topics}
To investigate the topical differences between catalyst and matched posts, we applied Empath~\cite{fast2016empath} to the posts' text. Empath scans each post against 200 topical lexicons and counts the number of matched words for each topic. In this analysis, we focused on a narrower set of 38 topics that fall in the following eight categories: affect, positive, negative, leisure \& travel, family \& friends, food, religion, and other.

Figure~\ref{fig:topics} shows topics covered in catalyst and matched posts. Both groups are more likely to include positive topics, such as  celebration, optimism, and friends; and less likely to include negative topics, such as disgust, rage, and anger. Positive topics are even more likely to occur in catalyst vs. matched posts, e.g., words related to positive topics, celebration, and optimism are more common in catalyst posts. On the other hand, topics related to leisure \& travel (e.g., vacation, eating, beach, cooking) are less common in catalyst posts.

In summary, catalyst posts are longer and more likely to contain question marks. There is a higher chance that they are text-only posts, and that they are original content as opposed to a reshare of another post. Finally, catalyst posts cover more positive topics compared to matched~posts.

\section{User Catalystness\label{sec:user_cat}}
Catalyst comments are far from evenly distributed. We consider the 1.1M users who had 30 or more catalyst comments in the 4-week period. They constitute just 1\% of all users in the US who posted at least one post during the 4-week study period, but are responsible for almost a third of all catalyst comments. Thus, it is important to understand these users and their role on the platform~better.

Similar to the post-level analysis, we want to account for the differences between catalysts and other users stemming from their activity level. To this end, we first match each catalyst to other people who posted the same number of posts, and then among them we pick the person with the closest number of friends to the catalyst. In all cases, we found an exact match in terms of number of posts. The average absolute difference in number of friends was 6.8, which is small given the fact that these users have many friends. We decided to match on the number of posts and the number of friends since users who post more often have more opportunities to catalyze interaction among their friends, and users with more friends have a higher number of potential interactions they can catalyze. We do not match users on the number of comments received as we are interested in comparing social catalysts with all other users rather than comparing social catalysts with just influencers. Also, we do not match users on other characteristics, such as their demographics or network position, as we are interested in how these characteristics vary between catalyst and non-catalyst users. In the rest of this section, we compare catalyst users with matched users who post as frequently, and have a similar number of friends.

\subsection{Demographics}
First, we compare basic demographics and find that on average, catalysts are 2.82 (95\% CIs [2.74, 2.91]) years older than matched users and have slightly longer tenure on Facebook by 4.72 (95\% CIs [4.49, 4.94]) months. Catalysts are also 11.7\% (95\% CIs [11.5\%, 11.9\%]) more likely to be male.

\subsection{Network structure} \label{subsec:network-structure}
The most related stream of work to our study of social catalysts is the research on social capital (Section~\ref{sec:related_work}). The main argument of that work is that an individual's position in their social network gives them a unique advantage, e.g., allowing them to control the flow of information~\cite{burt2005brokerage,burt2009structural,burt2017structural} or helping the enforcement of social norms~\cite{coleman1988social}, and increases their social capital.

In this section, we compare the networks of catalyst versus matched users to investigate the relationship between catalystness and network structure. We focus on the structural characteristics of the users' ego networks, i.e., the friendship edges between their friends, and consider a 10\% sample of all catalyst and matched users. As an illustration, Figure~\ref{fig:survey_net_viz} shows two example ego networks of a survey respondent and a friend the respondent nominated as catalyst (we describe the survey in more detail in Section~\ref{sec:survey}).

Given a user $i$ and their ego network $G$, we compute the following network characteristics:
\begin{description}
\item[\textbf{Density}] is the fraction of edges observed in $i$'s ego network over the number of possible edges, given the observed number of nodes.

\item[\textbf{Degree average}] is another measure of how likely $i$'s friends are to know each other.

\item[\textbf{Degree variance}] measures whether some nodes are more connected to other friends of the ego or they all have a similar number of connections to other nodes.

\item[\textbf{Degree assortativity}] measures the extent to which nodes preferentially connect to other nodes with similar degree. Assortative networks, in which high-degree nodes tend to connect to other high-degree nodes, have a core/periphery structure in which high-degree nodes clamp together to form the core and are surrounded by a less dense periphery of nodes with lower degree~\cite{newman2010networks}.

\item[\textbf{Average clustering coefficient}] is the average of the local clustering coefficients of the nodes within $i$'s ego network~\cite{watts1998collective}. Given $i$'s ego network $G$, the local clustering coefficient of a node $v \in G$ is defined as the fraction of pairs of $v$'s neighbors that are connected out of all pairs of neighbors of $v$. The local clustering coefficient can also be thought of as a local version of betweeness centrality, which measures the extent to which a node lies on (shortest) paths between other nodes.

\item[\textbf{Algebraic connectivity}] measures how easily information can flow within the network. Intuitively, it quantifies how likely it is for a random walk started in one part of the network to end up in another part of the network. More formally, it is the second eigenvalue of the Laplacian matrix of the graph~\cite{fiedler1973algebraic}.

\item[\textbf{Modularity}]measures the density of connections within communities vs. connections across communities~\cite{newman2004finding}. To find communities within the ego networks, we apply the algorithm proposed by Clauset, Newman, and Moore~\cite{clauset2004finding}, which uses greedy optimization to detect communities with maximum modularity.

\item[\textbf{Number of connected components in}]\textbf{the $k$-core and $k$-brace} of $i$'s ego network measures the structural diversity of $i$'s network. The intuition is that each connected component represents a different social context and that a higher number of connected components suggest higher social context diversity. The structural diversity of one's ego network on Facebook has been shown to be highly predictive of one's engagement on the platform~\cite{ugander2012structural}. The $k$-core of a graph is the subgraph of nodes obtained by repeatedly deleting nodes with degree less than $k$, and the $k$-brace of a graph is the subgraph formed by repeatedly deleting all edges of embeddedness (i.e., number of common neighbors shared by the two endpoints) less than $k$. The reason for analyzing the $k$-core/$k$-brace of the network, rather than the full network, is that most nodes are part of the network's giant connected component~\cite{newman2001random} and, thus, pruning the network helps us see more structure.
\end{description}

\begin{figure}[t]
\centering
\includegraphics[width=\linewidth]{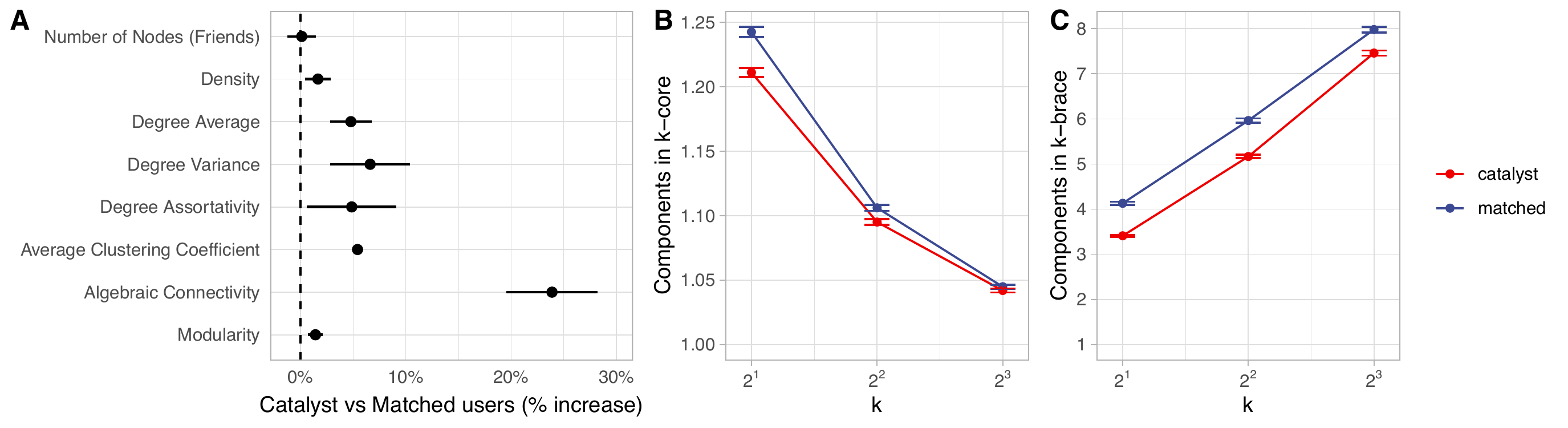}
\caption{Comparison of the structural characteristics of the ego networks (i.e., the network of connections among user's friends) of catalyst vs. matched users. The error bars represent 95\% confidence intervals.}
\label{fig:ego_cc}
\end{figure}

The results are shown in Figure~\ref{fig:ego_cc}. We find that, on average, due to matching, catalysts and matched users have an non-significant difference in the number of nodes in their networks, i.e., friends (95\% CIs [-1.22\%, 1.5\%]).

We note that the density of $i$'s ego network is equal to $i$'s local clustering coefficient in the full Facebook social network. This equivalence is particularly important since it relates to Burt's notion of structural holes\footnote{In Burt's original studies he used another network measure called redundancy. Subsequent work~\cite{borgatti1997structural} has shown that the local clustering coefficient can be thought of as simply a version of redundancy rescaled to have a maximum value of 1.}. According to Burt~\shortcite{burt2005brokerage,burt2009structural,burt2017structural}, the absence of connections among $i$'s friends gives $i$ power over the information flow between its friends and increases $i$'s social capital. If catalyst behavior can be explained by Burt's notion of social capital, we would expect social catalysts to be surrounded by structural holes, i.e., more likely to have less dense ego networks. On the other hand, if $i$'s friends are more acquainted with each other and know that the catalyst and their other mutual friends are observing their interactions, they might be more likely to interact, suggesting that catalysts should have denser ego networks. This aligns with Coleman's view of social capital~\cite{coleman1988social}. He argues that network closure allows for the enforcement of social norms and increases social capital. We do not find strong evidence for either of these two hypotheses: catalysts have only 1.7\% (95\% CIs [0.45\%, 2.9\%]) denser networks than the matched users, which translates to less than 0.001 difference in the average density of the two groups, 0.05707 vs. 0.05613.

Catalysts' ego networks also have higher degree average by 4.2 (4.8\% increase, 95\% CIs [2.83\%, 6.77\%]) and higher degree variance (6.6\% increase, 95\% CIs [2.85\%, 10.4\%]), suggesting that their networks are denser and that there is more diversity in how acquainted their friends are with each other.

Next, we look at how the edges in the users' ego networks are distributed, i.e., whether they are randomly dispersed across the network or clumped together among groups of friends. We find that catalysts' ego networks have 4.9\% (95\% CIs [0.63\%, 9.13\%]) higher degree assortativity or 0.0027 in absolute terms (0.0573 vs. 0.0546), indicating that their gregarious friends are slightly more likely to know each other. Catalysts' ego networks have 5.4\% (95\% CIs [4.95\%, 5.91\%]) higher average clustering coefficient than matched users (0.411 vs. 0.390), hinting that there is more transitivity among their friends' connections. Their networks also have higher algebraic connectivity suggesting that information can flow more easily among their friends. Compared to matched users, catalysts' ego networks have slightly higher modularity (0.3680 vs. 0.3627, i.e., 1.4\% increase, 95\% CIs [0.75\%, 2.13\%]), showing that both groups have similar level of community structure in their networks. When we consider the $k$-core (Figure~\ref{fig:ego_cc}B) and $k$-brace (Figure~\ref{fig:ego_cc}C) subgraphs, we find that, across all values of $k$, catalysts have fewer connected components in their ego networks, again suggesting that their social circles are tighter.

In summary, we find only small differences between the ego networks of catalysts and matched users across many structural measures. We note that these differences are statistically significant but substantively small. This suggests that catalyst behavior cannot be simply explained by the catalysts' network position.

\begin{figure}[t]
\centering
\includegraphics[width=0.5\linewidth]{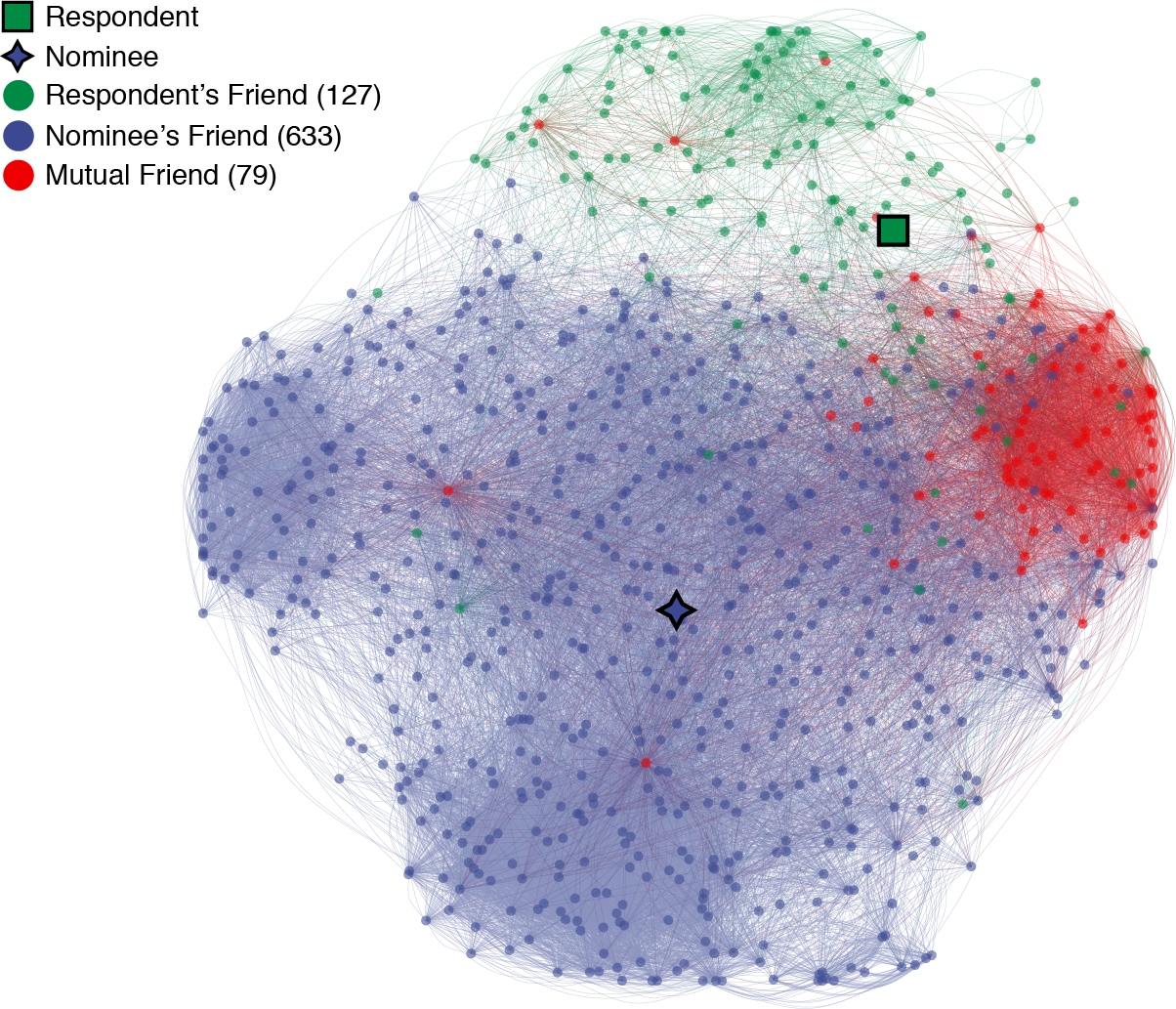}
\caption{Example of ego networks and their overlap (red) for a survey respondent (green) and their nominated friend (blue).}
\label{fig:survey_net_viz}
\end{figure}

\subsection{Friendship initiations and transitivity \label{subsec:friendship-initiations}}
In this section, we compare the friendship initiation activity of catalyst versus matched users to investigate whether catalysts, in addition to prompting discussion, also contribute to new friendship formation. Contrary to expectation, we find that catalysts are not more likely to be the initiator of new Facebook friendships. We find that, in aggregate, social catalysts have initiated 45.9\% of their friendships, whereas matched users have initiated 50.9\% (95\% CIs $<$0.01). The lower friendship initiation rate of catalysts is surprising, but some of the differences could be due to catalysts having longer tenure on Facebook compared to matched users and the reverse correlation between tenure and friendship initiation (users typically add many friends when they join Facebook). Moreover, if catalysts produce more engaging content, then it is likely that more people are interested in being their friends; however, confirming this hypothesis requires further analyses.

Next, we analyze the triadic closures that catalyst and matched users are part of to investigate the role that catalysts might be playing in connecting their friends. For each triadic closure, we compute the fraction of cases where the link between the friends of the catalyst or the matched user was formed after the links to their friends. In other words, if A is a catalyst or a matched user and an A---B---C triangle exists, we compute the fraction of cases where B---C was formed after both A---B and A---C. If such cases occur more than one-third of the time for catalysts, then that suggests that catalysts are more likely to be connecting their friends with each other. Indeed, we find that in 34.25\% of all triadic closures that catalysts are part of, the link between their friends was created last. On the other hand, for matched users, this number is 33.36\%, almost equal to the expected value of one-third. To conduct these analyses, we constructed a network of 11.6B links and note that the reported differences are statistically significant (95\% CIs $<$0.001).

\section{Survey: Perception of Catalyst Activity}
\label{sec:survey}
So far, we have used our measure of catalystness (Section~\ref{sec:measuring-catalystness}) to guide our analysis of the characteristics of catalyst users and posts. While the proposed definition of catalystness is the most direct measure of what we consider catalyst behavior, it may not necessarily be aligned with users' perceptions of catalystness. To this end, we conduct a survey in which we ask users to nominate friends that they consider to be catalysts. The survey is crucial in grounding our study of catalysts as it allows us to ($i$) test whether our measure of catalystness matches what users consider to be catalyst behaviors, and ($ii$) investigate whether the patterns we have found so far are robust and align with the user nominations.

\subsection{Survey design}
We invited a random sample of English speaking users in the US to participate in the survey. They were shown a banner on the top of their news feed inviting them to answer a short survey. After clicking the banner the users were informed that they will be asked questions about their experience on Facebook and that their answers will be kept confidential.

We asked users the following four questions:
\begin{itemize}
    \item \textbf{Q1:} Who posts things on Facebook that have a lot of entertaining replies and comments?
    \item \textbf{Q2:} Who posts things on Facebook that a lot of people like and enjoy?
    \item \textbf{Q3:} Who posts things on Facebook that get people to open up and talk with others?
    \item \textbf{Q4:} Who posts things on Facebook that get you to interact with others?
\end{itemize}

The questions were designed to capture different aspects of catalyst behavior. Each question was followed by the following statement: ``Please list up to 3 Facebook friends that best fit this description'', and a field to input the names of the nominated users. To reduce the cognitive load on the participants, we provided a typeahead functionality: as the users started typing the name of the nominee a dropdown list with the profile pictures and names of the users that match the prefix appeared\footnote{This user interaction is present in many other parts of Facebook and is familiar to the users.}. Here we enumerated the question as Q1 to Q4 for reference, but in the survey, both the questions and the typeahead suggestions were presented in a random order to avoid any ordering effects. The questions were shown one at a time and the respondents did not know what they will be asked next. None of the questions were required and the respondents could skip a question at any point of the survey.

We ran the survey for a week and collected 10,351 responses with at least one question answered. Among the people who participated (i.e., answered at least one question) 44.17\% answered all four questions, and all questions had similar response rates, ranging between 48\% to 53\%. To check for any ordering effects, we reran all subsequently reported analyses using only the responses to the question when it was shown first and did not find any substantive differences.

\subsection{Survey sample}
While we invited a random sample of users to participate in the survey, the users who responded, compared to the population of active Facebook users, were 1 year older, were 11\% more likely to be female, had 182 more friends, and had been using Facebook for 9.5 months longer. We also find homophily in the respondents' nominations: 69.2\% of all nominations by women were women, and 55.5\% of all nominations by men were men.

\begin{figure}[t]
\centering
\includegraphics[width=0.5\linewidth]{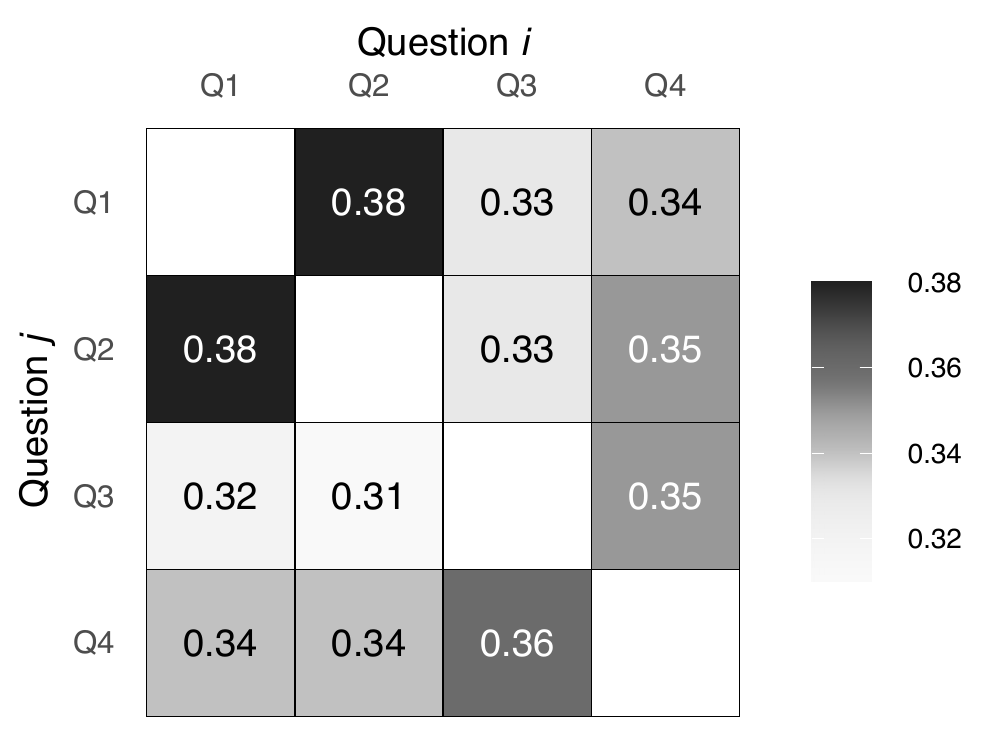}
\caption{Overlap between the users nominated in each of the four survey questions. The cell values show the probability of a user being nominated in question $i$, given that they have been nominated in question $j$. Note that the questions were presented in random order.}
\label{fig:survey_overlap}
\end{figure}

\subsection{Number and overlap of nominations} 
As mentioned above, we asked respondents to nominate up to three friends that fit the description in the question the best. Given the high cognitive load of the question, we expected a low number of nominations per question. To our surprise, the majority of the respondents, across all four questions, nominated three friends (Q1: 71.8\%, Q2: 74.5\%, Q3: 71.2\%, Q4: 74.5\%). Our instructions may have had an anchoring effect, giving respondents the impression that three nominations were expected.

Next, we look at the overlap in the nominations between the four questions. Figure~\ref{fig:survey_overlap} shows the probability of the person who was nominated in one question to be also nominated in another one. This probability ranges between 0.31 and 0.38; Q1\&Q2 and Q3\&Q4 are more correlated with each other compared to other pairs, suggesting that posting things that prompt people to interact with others is somewhat distinct from making posts that are simply engaging.

\subsection{Nominated vs. other friends}
\begin{figure}[t]
\centering
\includegraphics[width=\linewidth]{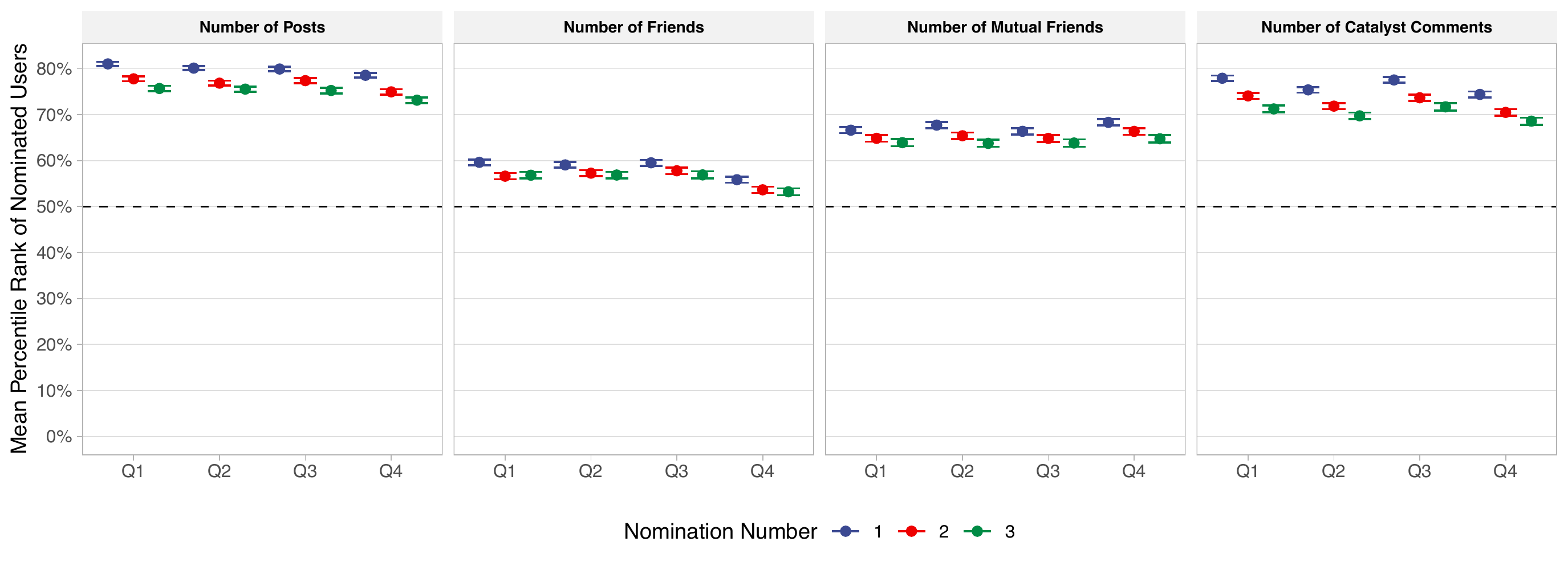}
\caption{Comparison between the nominated and all other respondent's friends in terms of their activity on Facebook (number of posts), degree (number of friends), strength of the tie with the respondent (number of mutual friends), and catalystness (number of total catalyst comments prompted by the user). The error bars represent 95\% confidence intervals.}
\label{fig:survey_nom_percentiles}
\end{figure}

We compare the characteristics of the nominated to the characteristics of the other respondents' friends. We consider four characteristics: activity level (number of posts), degree (number of friends), tie strength (number of mutual friends), and catalystness (total number of catalyst comments on the users' posts over the previous 28 days). For each characteristic, we rank all respondents' friends and check where the nominated friends fall. We randomly order friends with the same value. Figure~\ref{fig:survey_nom_percentiles} shows the percentile rank of the nominated users for each of the four characteristics. If the characteristic did not play a role in the nomination process, i.e., respondents chose at random, then we would expect the percentile rank to be close to 50\% (represented as a dashed line in Figure~\ref{fig:survey_nom_percentiles}).

We find that the nominated friends post more than the other respondents' friends, they fall within 73.1 and 81 percentile rank. While the nominated users post more frequently than many of the respondents' friends, they are far from \textit{superposters}~\cite{graham2014discursive}, i.e., among the top 1\% most frequent posters. The friends' degree is less important, the nominated friends fall between the 53.2 and 59.6 percentile rank. This suggests that the nominated users do not fit the definition of influencers, who tend to have a large number of friends. However, the nominated friends tend to be closer to the respondent than a randomly chosen friend, but they are not the respondents' closest friends, falling between the 63.8 and 68.3 percentile rank. Finally, the nominated friends get more catalyst comments than other respondents' friends---68.6 to 77.9 percentile rank---providing evidence that our measure of catalystness correlates with the respondents' perceptions.

We also observe a clear and consistent pattern suggesting that the order of the nominations matters. The users nominated first have higher percentile rank across all questions and all characteristics.

\subsection{Matching}
In the previous subsection, we showed that the nominated friends have a higher number of catalyst comments than other respondents' friends. However, we looked at each of the four characteristics individually and it might be that the number of catalysts comments is simply a proxy for other user characteristics (e.g., activity level).

To rule out this hypothesis, we perform a matching analysis similar to that in Sections~\ref{sec:post-catalystness} and \ref{sec:user_cat}. We match each nominated friend to another, most similar, friend of the respondent. Since we know the relative importance of each characteristic, we matched users sequentially: we first narrow down the matches by number of posts, then by number of mutual friends (rounded to the closest 50), and finally by number of friends.

\begin{figure}[t]
\centering
\includegraphics[width=0.55\linewidth]{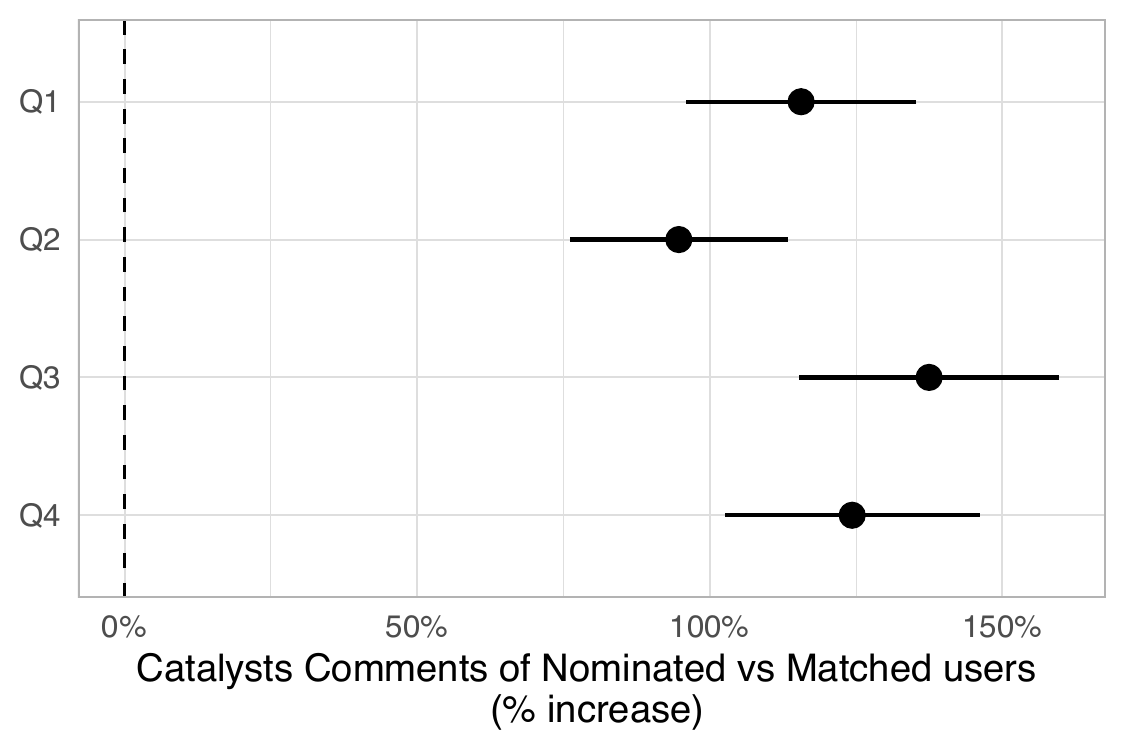}
\caption{Comparison between the nominated and matched users by number of catalyst comments on their posts. The points represent percent increase and the error bars represent 95\% confidence intervals.}
\label{fig:survey_catalystness_ratios}
\end{figure}

Given the matched pairs we compare the percent increase in the number of catalyst comments of the nominated vs. the matched users (Figure~\ref{fig:survey_catalystness_ratios}). We find that across all questions the nominated users have 94.7\% to 137\% more catalyst comments than their matched counterparts. Question 3, which aligns most closely with our definition of catalystness, has the highest increase, 137\% (95\% CIs [115\%, 160\%]).

The matching analysis, together with the overall comparison of the nominated vs. all other respondents' friends in the previous subsection, provides strong evidence that the proposed measure of catalystness aligns well with the respondents' perceptions of catalystness.

\subsection{Characteristics of nominated vs. matched users}
Beyond allowing us to test the alignment of our catalystness measure with users' perceptions, the survey also allows us to test the robustness of the patterns of catalyst user characteristics we found in Section~\ref{sec:user_cat}. To do so, we repeated all earlier analyses, but instead of using the catalystness measure to determine which users are catalysts, we used the respondents' nominations and compared their characteristics to the matched respondents' friends.

We find very similar patterns to those reported in Section~\ref{sec:user_cat} across all four questions. One exception is the gender breakdown of the catalysts: here we find that men are not more likely to be catalysts and that the same fraction of catalysts and matched users are men or women. This discrepancy could be explained by the composition of our survey sample, i.e., 1.85 times more women, and the homophily in the respondents' nominations, i.e., women are 2.25 times more likely to nominate other women.

\section{Modeling Catalystness}
\label{modeling-catalystness}
While we showed that the catalystness score is a good proxy of users' perceptions, it is not a perfect measure. In this section, we build a predictive model that distinguishes between nominated users and other respondents' friends. This allows us to ($i$) test how accurately we can model users' perceptions and ($ii$) compare the predictive power of different measures at the same time.

We focus on Q3 and all the 19k nominated users. Using data from a four week period before the survey was conducted, we extract the following features, most of which have been described in earlier sections:

\begin{itemize}
    \item \textbf{Catalyst comments:} overall and by the relationship type of the catalyzed users (friends, friends of friends, other).
    \item \textbf{Demographics:} age, gender, friend count, etc.
    \item \textbf{Activity level:} number of posts, English posts, text posts.
    \item \textbf{Post characteristics:} length, content type, frequency of mentions, emojis, hashtags, etc.
    \item \textbf{Post topics:} eight topical categories: affect, positive, negative, leisure \& travel, family \& friends, food, religion (full list of topics shown in Figure~\ref{fig:topics}).
    \item \textbf{Post sentiment:} positive, negative, neutral, and compound score measured using Vader~\cite{hutto2014vader}.
    \item \textbf{Reactions:} to user's posts and their comments.
    \item \textbf{Groups activity:} membership, administratorship, passive and active engagement measures.
    \item \textbf{Event participation:} number of events created, joined, as well as number of invites sent and received, etc.
    \item \textbf{Friend-respondent tie strength:} number of mutual friends.
\end{itemize}

Overall, we have 118 features. We split the data 80-20\% for training and testing and balance the training set by under-sampling negative labels. Then, we train a Gradient Boosting Decision Trees (GBDT) model, which has been shown to achieve state-of-art results in many classification tasks~\cite{friedman2001greedy}. To prevent over-fitting, we limit the number of iterations to 100, the depth of the trees to five, and the number of leaves to 20. For this classification task we obtain a fairly high AUC of 78.85\%, precision of 55.37\%, recall of 95.72\%, and F1 score of 0.7016.

A benefit of using GBDT is that after the boosted trees are constructed, we can rank and compare the relative importance of each feature. Intuitively, this importance is computed by considering how much the performance was improved in each decision tree for a given split. This is then weighted by the number of decisions that were made through that node and finally averaged across all trees. Table~\ref{tab:top_features} shows the top five most important features according to our model, showing that catalyst comments per post among friends is one of the top five features among the many features we considered.

As we discussed in Section~\ref{sec:survey} the order of friend nominations matters. We find that if we restrict the data to the first nominated user by each respondent, we obtain a substantially more accurate model, with AUC of 82.44\%.

\begin{table}[t]
\caption{Top five most important features in our Gradient Boosting Decision Trees model that distinguishes between the respondents' friends nominated as catalysts and all other friends.}
\begin{center}
\begin{tabular}{lc}
\toprule
Feature & Importance \\
\midrule
Number of friends  & 8.9\% \\
Number of comments on posts  &  5.7\% \\
Number of posts  &  5.6\% \\
Number of catalyst comments/post among friends  &  4.9\% \\
Number of \textit{hahas} on comments  & 4.1\% \\
\bottomrule
\end{tabular}
\end{center}
\label{tab:top_features}
\end{table}

\section{Discussion and Conclusion}
\label{sec:discussion_conclusion}

\subsection{Summary}
In this paper, we studied social catalysts, people who are good at bringing others together, both using observational data and a survey. We defined a catalystness measure as the number of comment-reply pairs on users' posts that do not include the poster. Then, we compared posts with high catalystness scores with posts by similar users and found that catalyst posts are longer, more likely to be original content than reshares, and more likely to include positive topics compared to the matched, non-catalyst posts.

Using the catalystness scores of the posts, we defined social catalysts as users whose posts have high catalystness scores. Then, we compared these users with other users with similar activity levels and number of friends. We found that catalysts are more likely to be men and are slightly older. Surprisingly, we also found that there are only small differences in the structure of the ego networks of the catalysts and matched users.

Through a survey of 10.3k users, we showed that the proposed measure of catalystness is well-aligned with the respondents' perceptions of catalystness. We also trained a model that distinguishes between the nominated and other friends of the respondent to compare the importance of different user characteristics and to measures how well we can model the respondents' perceptions. 
We found that we can model the survey participants' nominations with fairly high accuracy and that the users' activity levels and catalystness scores are among the most predictive features. The ability to model catalyst activity could be helpful in ranking and recommendations, creating more opportunities for people to interact online.

\subsection{Applicability to other social media platforms}
In this study, we focused on posts and users that catalyze interactions on Facebook, but our measure of catalystness can be readily applied to other social media platforms that allow users to comment and reply to posts by other users. As a reminder, we defined the catalystness score of a post as the number of interactions among users other than the poster, and the catalystness of a user as the sum of all catalyst interactions prompted by the user's posts (Section~\ref{sec:measuring-catalystness}). The only requirement for computing this measure is to be able to infer who replied to whom, in order to exclude replies from and to the poster. This can be achieved by first constructing a user reply graph, similar to the one depicted in Figure 1. Depending on the specific social media platform, such graphs can be constructed by either relying on user mentions (available on most platforms) or by taking into account the explicit structure of the conversation, e.g., the hierarchical nesting of the comments on platforms akin to forums like Reddit. Many previous studies have relied on constructing user reply graphs to analyze the interaction patterns on various platforms, including Twitter~\cite{saveski2021structure, coletto2017automatic}, YouTube~\cite{thelwall2014analysing, shoham2013writing}, Wikipedia~\cite{welser2011finding, iosub2014emotions}, Reddit~\cite{hessel2019something, buntain2014identifying}, Slashdot~\cite{gomez2008statistical, gonzalez2010structure}, and Usenet~\cite{fisher2006you, gleave2009conceptual}. However, we note that studying social catalysts on Facebook allowed us to analyze the relationship between catalystness and the users' social network structure.

\subsection{Practical implications}
The goal of our work was to characterize and understand social catalyst activity in an online social network. Our measure of catalystness is a tool that researchers and social media platforms alike can use to characterize a new aspect of online conversations as well as an understudied role in social networks, online and offline. Having this measure may encourage further research into characteristics and behaviors that prompt interactions. 

Beyond this, the findings about social catalysts have several important practical implications. First, being able to identify catalyst activity can help platforms understand their prevalence in different contexts. These interactions may be beneficial both for the health of the platform and the well-being of the users, as active user engagement is associated with increased well-being~\cite{burke2016wellbeing}. 
Second, as discussed in Section~\ref{sec:user_cat}, since a small proportion of users are responsible for a disproportionate amount of catalyzing activity, they provide value and positively affect the experiences of many other users on the platform. The methodology proposed in this study could be used to further understand how this activity evolves and how it can best be supported. 
Third, our measure of catalystness and predictions of whether a post or a user is a catalyst could potentially be used as signals in friend recommendations or post ranking models. However, whether connecting to catalysts or seeing catalyzing posts could have the expected positive effects, such as helping new users integrate or strengthening the social fabric among existing users or whether it might have other effects, would require further study.

\subsection{Limitations}
Our study has several limitations. First, throughout the paper, we relied on observational data---behavioral logs and survey responses---to study social catalysts and thus our analyses are correlational rather than causal. We believe, however, that the findings in this work can guide the choice of causal questions for further study.

Second, we used data from one month and a single snapshot of the users' social networks to study the characteristics and behaviors of social catalysts. A more longitudinal view of users' behavior will allow us to study whether being a social catalyst is a transient characteristic and how catalystness varies over time. The strong alignment of our results with the perceptions of the survey respondents suggests that our patterns are robust; the respondents, presumably, form their impression of who among their friends is a catalyst by observing their behaviors over a longer period.

Moreover, we focused only on users in the US. This allowed us to perform linguistic analysis on English-language text and minimized potential confounding effects due to different social norms across countries~\cite{yuki2005cross}. However, we note that the patterns of user engagement may vary internationally based on the levels of adoption of Facebook, cultural differences, and social norms. For instance, previous work studying communication on Twitter has found that networks of users in Korea, Japan, and Singapore have higher and more variable local clustering coefficients than users in the US and the UK~\cite{park2018strength}. This suggests that social catalysts may be more common in some countries than in others.

Finally, due to privacy constraints, we are unable to share any user-related data. However, to ease the reproducibility of our work, we restricted our analysis to metrics that are publicly available on other platforms, omitting metrics such as the number of post views.

\subsection{Future work}
One of the key open questions is what the impact of social catalysts is on other users. For instance, if a new user joins Facebook and becomes friends with a catalyst early on, would they have a more engaging experience on Facebook than they would if they friended a non-catalyst? Similarly, when catalysts are less active or leave the platform altogether, do they leave a void behind? Do their friends interact on other people's posts or does the amount of their interaction decrease?

\section{Acknowledgments}
We would like to thank Justin Cheng, Bogdan State, Ismail Onur Filiz, Guanghua Chi, Johan Ugander, and the anonymous reviewers for their helpful comments. We are also thankful to the editors for their patience.

%
%
\bibliographystyle{ACM-Reference-Format}
\bibliography{refs_clean}

\received{June 2020}
\received[revised]{October 2020}
\received[accepted]{December 2020.}

\end{document}